\documentclass[3p,times,twocolumn]{elsarticle}

\usepackage{ecrc}

\usepackage{hyperref}

\usepackage[utf8x]{inputenc}
\usepackage{amsmath}
\usepackage[caption=false]{subfig}

\newcommand{\der}{\mathrm{d}}
\newcommand{\rt}{{\mathbf{r}_T}}
\newcommand{\xt}{{\mathbf{x}_T}}
\newcommand{\bt}{{\mathbf{b}_T}}
\newcommand{\bti}{{\mathbf{b}_{T,i}}}

\newcommand{\yt}{{\mathbf{y}_T}}

\newcommand{\ud}{\, \mathrm{d}}

\newcommand{\tr}{\, \mathrm{Tr} \, }

\newcommand{\nc}{{N_\mathrm{c}}}

\newcommand{\gev}{\ \textrm{GeV}}

\newcommand{\sigmap}{{ \sigma^\textrm{p}_\textrm{dip} }}

\newcommand{\dsigmap}{{\frac{\ud \sigma^\textrm{p}_\textrm{dip}}{\ud^2 \bt}}}

\newcommand{\xpom}{{x_\mathbb{P}}}

\volume{00}

\firstpage{1}

\journalname{Nuclear and Particle Physics Proceedings}

\runauth{}

\jid{nppp}

\jnltitlelogo{Nuclear and Particle Physics Proceedings}

\usepackage{amssymb}

\usepackage[figuresright]{rotating}

\begin{document}

\begin{frontmatter}



\dochead{}

\title{Constraints for proton structure fluctuations from exclusive scattering}


\author{H. Mäntysaari and B. Schenke}

\address{Physics Department, Brookhaven National Laboratory, Upton, NY 11973, USA}

\begin{abstract}
We constrain the average density profile of the proton and the amount of event-by-event fluctuations by simultaneously calculating the coherent and incoherent exclusive diffractive vector meson production cross section in deep inelastic scattering. Working within the Color Glass Condensate picture, we find that the gluonic density of the proton must have large geometric fluctuations in order to describe the experimentally measured large incoherent cross section. 
\end{abstract}

\begin{keyword}
Diffraction \sep fluctuations \sep CGC \sep exclusive reaction


\end{keyword}

\end{frontmatter}


\section{Introduction}

The proton is a quantum mechanical object, whose \emph{average} structure in terms of quark and gluon densities is accurately measured in deep inelastic scattering experiments especially at HERA~\cite{Abramowicz:2015mha}. To get a more detailed picture of the proton structure, one is fundamentally interested in the Wigner distribution~\cite{Belitsky:2003nz} which would provide access to both transverse momentum and spatial distribution of the constituents. However, it is not known how to measure such a distribution. In the near future, the Electron ion Collider~\cite{Accardi:2012qut} will make it possible to investigate the transverse momentum dependent or generalized parton distribution functions (TMDs and GPDs) where some of the degrees of freedom are integrated out. In Refs.~\cite{Mantysaari:2016ykx,Mantysaari:2016jaz} we showed how exclusive diffractive vector meson production in deep inelastic scattering can be used to add information about the geometric structure fluctuations into this picture.

Revealing the details of the proton structure and the QCD dynamics responsible for it is an interesting task in itself. In addition, the interpretation of experimental measurements on proton-nucleus collisions requires detailed information about the initial state of the collisions. In particular, there are many collective phenomena observed in pA collisions that have been traditionally associated with hydrodynamical behavior  of heavy ion collisions (see e.g. Ref.~\cite{Dusling:2015gta} for a review). In order to study if pA collisions can also be described hydrodynamically, it is required to know the initial state geometry (and its event by event fluctuations), as the initial state geometry drives the collective phenomena in the final state. 

Diffractive vector meson production is a powerful tool to access the geometric structure of the proton. In such a process no color charge is exchanged with the proton target, and the experimental signature is a large rapidity gap between the produced vector meson and the proton or proton remnants. In the Good-Walker picture, the diffractive scattering is described in terms of states that diagonalize the scattering matrix~\cite{Good:1960ba}. At high energy
these states are the ones where a virtual photon fluctuates into a quark-antiquark color dipole, which interacts with a particular configuration of the target.

Averaging the cross section over different target configurations at the level of the cross section gives the total diffractive vector meson production cross section, where no restrictions are imposed on the final state of the proton. On the other hand, if the average is performed at the level of the scattering amplitude, the state of the target (the proton) is required to be the same before and after the interaction and the process is known as \emph{coherent diffraction}. Subtracting the coherent diffractive cross section from the total coherent cross section one obtains the cross section for the process where the proton is required to break up after the interaction, called \emph{incoherent diffraction}. As the incoherent diffractive cross section then becomes a variance, it is sensitive to fluctuations of the scattering amplitude. For more details, we refer the reader e.g. to Refs.~\cite{Miettinen:1978jb,Frankfurt:1993qi,Caldwell:2009ke}.

At high energy, a convenient framework to describe the scattering process is given by the Color Glass Condensate (CGC) effective theory of QCD. Coherent diffractive vector meson production measured by HERA has been successfully described within the CGC framework~\cite{Kowalski:2006hc}, as well as the fully inclusive DIS cross section~\cite{Albacete:2010sy,Lappi:2013zma,Rezaeian:2012ji}. The framework has also been applied to describe diffractive vector meson production off nuclei at an Electron Ion Collider~\cite{Lappi:2010dd} and to ultraperipheral collisions~ \cite{Lappi:2013am}.


\section{Diffractive vector meson production in the CGC framework}
Diffractive vector meson production can be naturally described in the Color Glass Condensate framework. First, an incoming virtual photon splits to a quark-antiquark dipole. This splitting is described by the light cone QED wave function of the photon $\Psi(Q^2, \rt, z)$. Here, $Q^2$ is the virtuality of the photon, $\rt$ the separation between the quark and the antiquark and $z$ the longitudinal momentum fraction of the photon carried by the quark. The dipole scatters off the target (in this work proton) without exchanging a net color charge. This scattering is given by the dipole cross section $\sigmap$. Finally, the vector meson is formed from the proton, described in terms of the virtual photon wave function $\Psi^V(Q^2,\rt,z)$. In this work, the Boosted-Gaussian wave function from Ref.~\cite{Kowalski:2006hc} is used. The diffractive scattering amplitude reads~\cite{Kowalski:2006hc}
\label{eq:diff_amp}
\begin{multline}
 A^{\gamma^* p \to J/\Psi p}_{T,L}(\xpom,Q^2, {\Delta}) = i\int \der^2 \rt \int \der^2 \bt \int \frac{\der z}{4\pi}  \\
 \times (\Psi^*\Psi_V)_{T,L}(Q^2, \rt,z) 
e^{-i[\bt - (1-z)\rt]\cdot {\Delta}}  \dsigmap(\bt,\rt,\xpom).
\end{multline}
Here $\bt$ is the impact parameter, $\xpom$ the longitudinal momentum fraction of the proton
 and the transverse momentum transfer is ${\Delta}$. The dipole-proton scattering amplitude is obtained from the IPsat  or IP-Glasma models discussed in Sec.~\ref{sec:dipole}.

The coherent cross section is obtained by averaging Eq.~\eqref{eq:diff_amp} over all the possible target proton configurations:
\begin{equation}
\label{eq:coherent}
\frac{\der \sigma^{\gamma^* p \to J/\Psi p}}{\der t} = \frac{1}{16\pi} \left| \langle A^{\gamma^* p \to J/\Psi p}(\xpom,Q^2,{\Delta}) \rangle \right|^2,
\end{equation}
with $t\approx -\Delta^2$.
When the target breaks up, the incoherent cross section is given by the variance of the scattering amplitude:
\begin{multline}\label{eq:incoherent}
\frac{\der \sigma^{\gamma^* p \to J/\Psi p^*}}{\der t} = \frac{1}{16\pi} \left( \left\langle \left|  A^{\gamma^* p \to J/\Psi p}(\xpom,Q^2,{\Delta})  \right|^2 \right\rangle \right. \\
- \left. \left| \langle A^{\gamma^* p \to J/\Psi p}(\xpom,Q^2,{\Delta}) \rangle \right|^2 \right)\,.
\end{multline}

\section{Dipole cross section}
\label{sec:dipole}
To describe the dipole-proton scattering we use two different models. The first one is the IPsat model, where the dipole-target cross section is written as
\begin{equation}\label{eq:unfactbt}
\dsigmap(\bt,\rt,\xpom)
 = 2\,\left[ 1 - \exp\left(-\rt^2  F(\xpom,\rt) T_p(\bt)\right) 
\right].
\end{equation} Here $T_p(\bt)$ is the proton spatial profile function which is assumed to be Gaussian, $  T_p(\bt)=\frac{1}{2\pi B_p} e^{-\bt^2/(2B_p)}$.
The function $F$ is proportional to the 
DGLAP evolved gluon distribution whose exact form can be found from Ref.~\cite{Rezaeian:2012ji}, where the free parameters
are obtained by performing a fit to HERA data.

The geometric fluctuations are included in this model by assuming that the small-$x$ gluons in the proton are localized around the constituent quarks (or three hot spots) at points $\bti$. The positions of the constituent quarks are sampled from a Gaussian distribution with width $B_{qc}$. Then, the proton density profile is assumed to be Gaussian around the centers of the constituent quarks, and the width of these distributions is denoted by $B_{q}$. 
The procedure amounts to the replacement
$T_p(\bt) \rightarrow \frac{1}{N_q} \sum_{i=1}^{N_q} T_q(\bt-\bti)$
in Eq.~\eqref{eq:unfactbt},
with $N_q=3$, and $T_q$ is the Gaussian of width $B_{q}$.

We also use the IP-Glasma framework that includes local color charge fluctuations and has been successfully used to describe the initial state of the heavy ion collisions. In this case the dipole-target cross section is obtained by first relating the color charge density locally to the saturation scale obtained from the IPsat model. Then, one solves the classical Yang-Mills equations and obtains the Wilson lines $V(\xt)$ at each transverse point. The dipole-target cross section is then obtained from the Wilson lines as $\dsigmap(\rt=\xt-\yt, \bt=(\xt+\yt)/2, \xpom) = 2(1 -  \tr [ V(\xt) V^\dagger(\yt) ] / \nc )$. The geometric fluctuations are included by using an IPsat model with geometric fluctuations. 
For more details of the IP-Glasma model, we refer the reader  to Ref.~\cite{Schenke:2012fw}.

\section{Results}
\begin{figure}[tb]
\centering
\vspace{-1em}
		\includegraphics[width=0.5\textwidth]{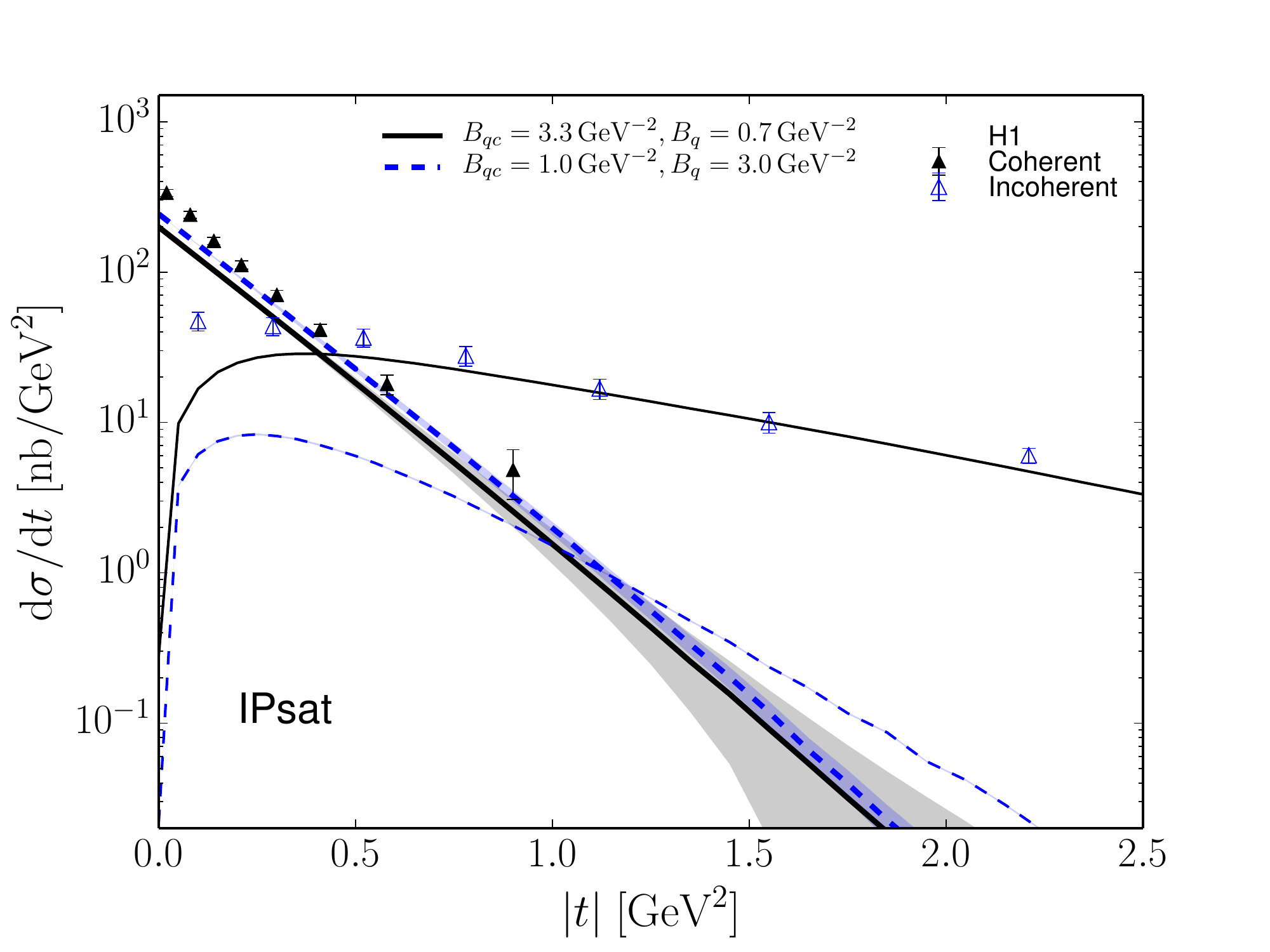} 
				\caption{Coherent (thick  lines) and incoherent(thin lines) diffractive $J/\Psi$ production at $\langle W \rangle = 75 \gev$ calculated with large (solid lines, see Fig.~\ref{fig:ipsat_protons}) and more moderate (dashed) geometric fluctuations using the IP-sat model. The results are compared with H1  data~\cite{Aaron:2009xp}. }
		\label{fig:ipsat}
\end{figure}

\begin{figure}[tb]
\centering
		\includegraphics[width=0.45\textwidth]{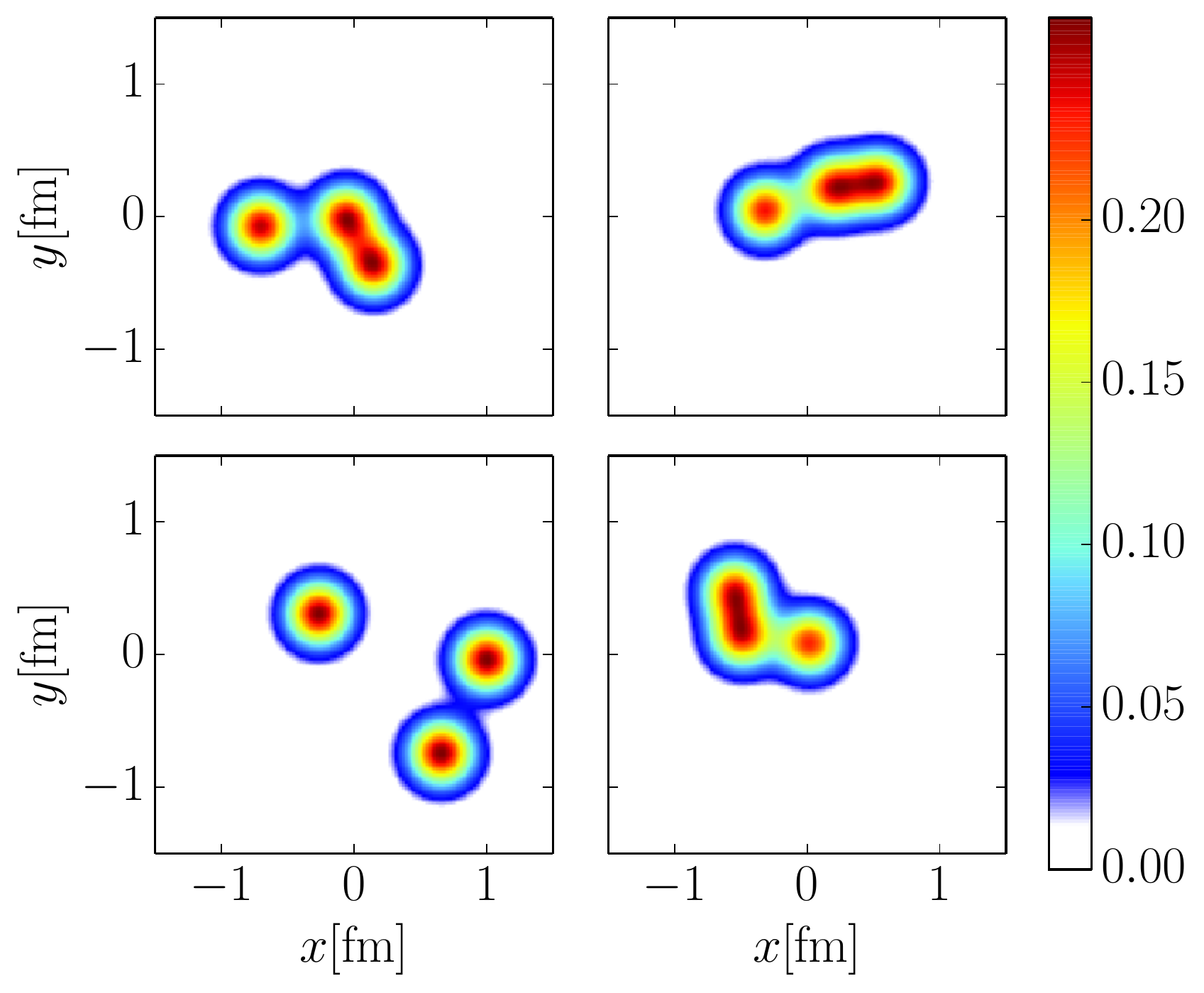} 
				\caption{Example density profiles of the protons with large geometric fluctuations. Parameters are $B_q=3.3\gev^{-2}$ and $B_{qc}=0.7\gev^{-2}$.  }
		\label{fig:ipsat_protons}
\end{figure}

Let us first present results obtained by using the IP-sat model.
The coherent and incoherent cross sections at $\langle W \rangle =75 \gev$ (corresponding to $\xpom \sim 10^{-3}$) are shown in Fig.~\ref{fig:ipsat}. A good agreement with the H1 data~\cite{Aaron:2009xp} is obtained when geometric fluctuations with parameters $B_{qc}=3.3\gev^{-2}, B_q=0.7\gev^{-2}$ are included. Example proton density profiles obtained from this parametrization are shown in Fig.~\ref{fig:ipsat_protons}. In Fig.~\ref{fig:ipsat} the cross sections are also calculated using a more smooth proton parametrization, which does not change the good agreement with the coherent cross section measurements, but significantly underestimates the incoherent cross section.

\begin{figure}[tb]
\vspace{-1em}
\centering
		\includegraphics[width=0.5\textwidth]{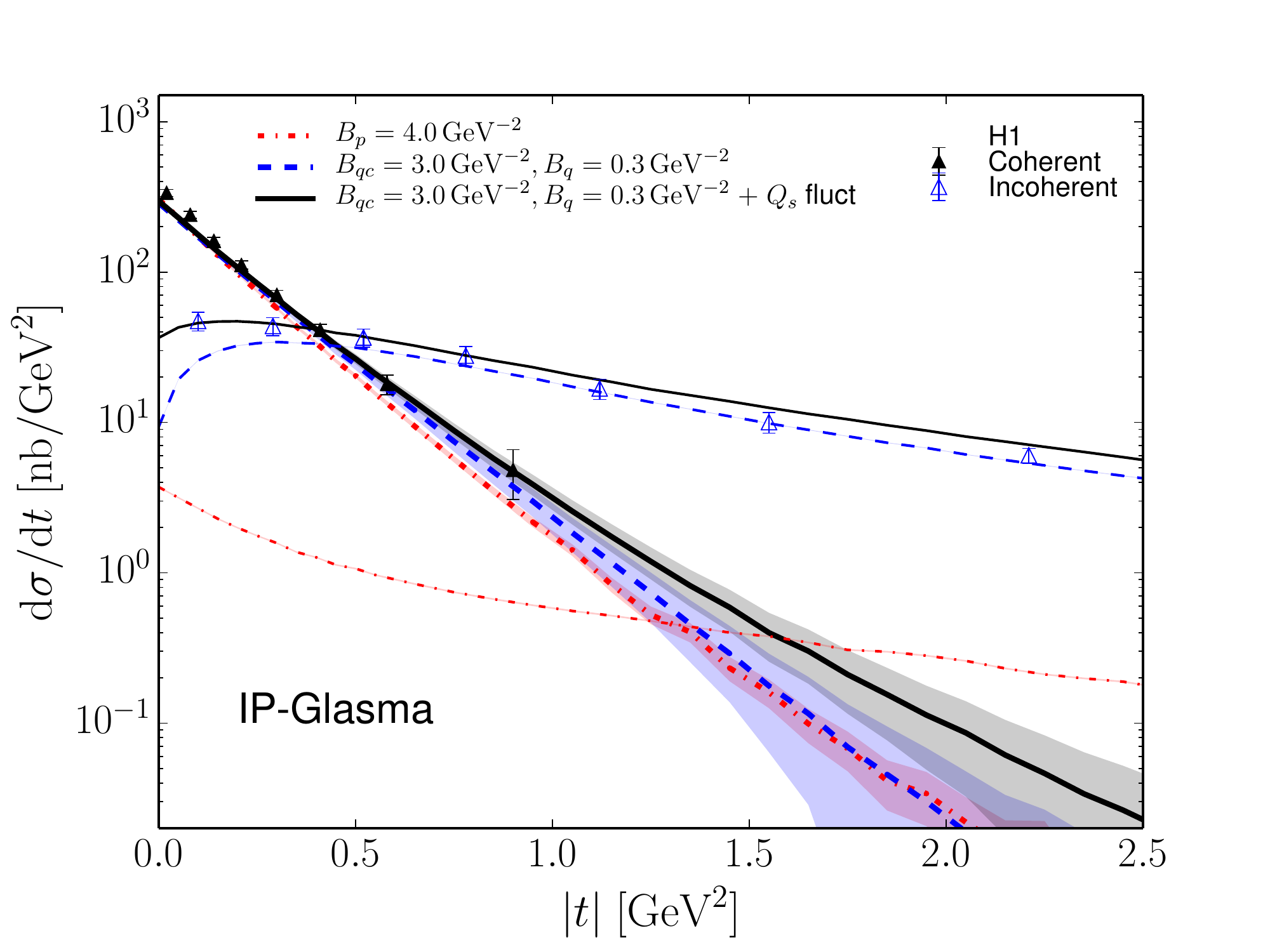} 
				\caption{Coherent (thick  lines) and incoherent (thin lines) diffractive $J/\Psi$ production at $\langle W \rangle = 75 \gev$ calculated with and without additional $Q_s$ fluctuations. For comparison, the result obtained for the proton with only color charge fluctuations is shown ($B_p=4\gev^{-2}$).  }
		\label{fig:ipglasma}
\end{figure}

\begin{figure}[tb]
\centering
		\includegraphics[width=0.45\textwidth]{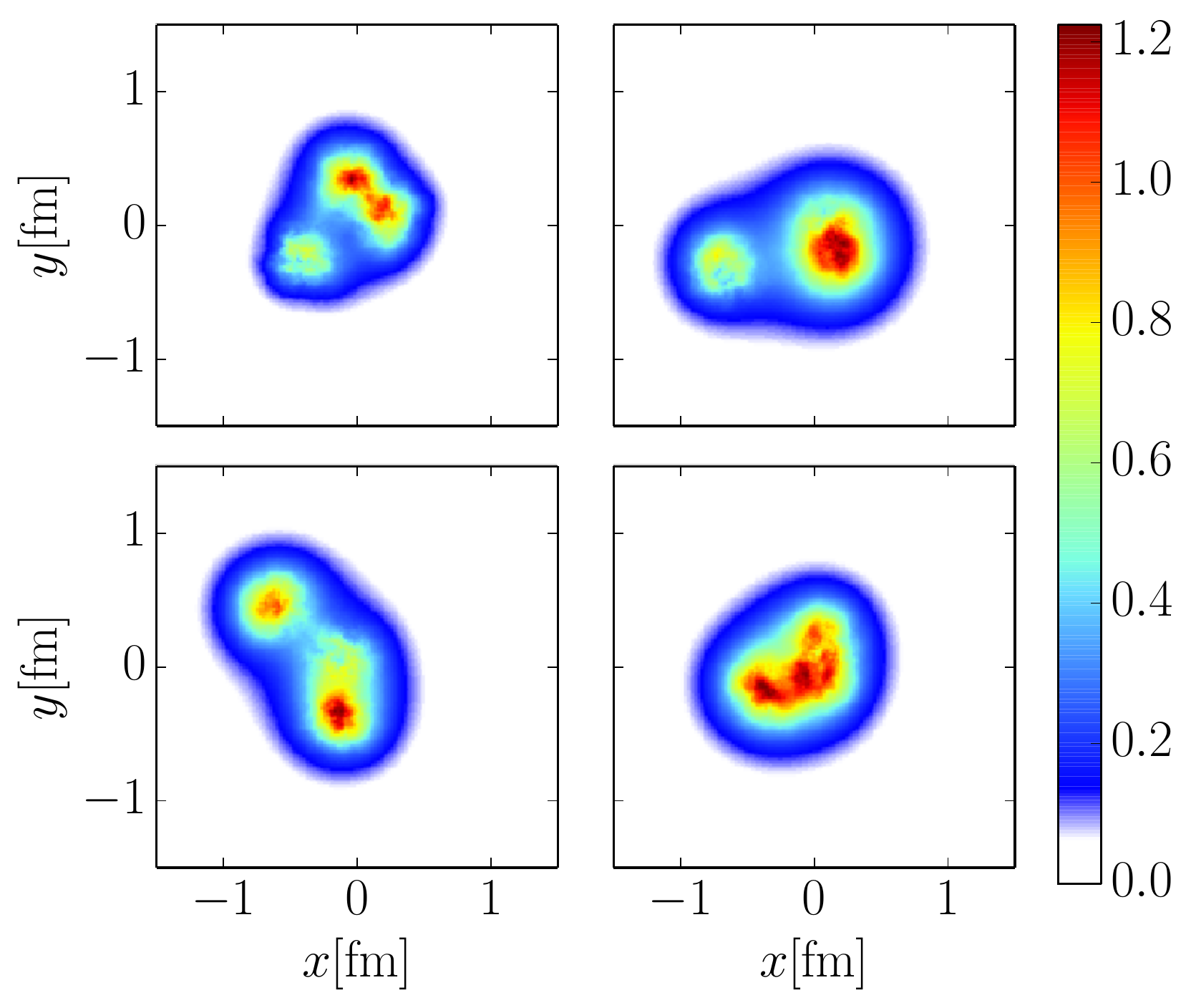} 
				\caption{Example proton density profiles ($1.0 -  \mathrm{Re} \tr V(x,y)/\nc$) with geometric fluctuations ($B_q=3.0\gev^{-2}$ and $B_{qc}=0.3\gev^{-2}$).  }
		\label{fig:ipglasma_protons}
\end{figure}

Similarly, the cross sections calculated using the IP-Glasma model are shown in Fig.~\ref{fig:ipglasma}. In this case, we show the results obtained without geometric fluctuations which is labeled as $B_p=4.0\gev^{-2}$.
Unlike in case of the IPsat model, one also obtains a nonzero incoherent cross section in that case because of the color charge fluctuations. However, we see that the color charge fluctuations alone are not enough to describe the measured incoherent cross section.
The geometric fluctuations are then included by again finding parameters $B_{qc}$ and $B_q$ that  give a good description of the H1 data. The amount of fluctuations in that case is demonstrated by showing in Fig.~\ref{fig:ipglasma_protons} a few example configurations.

Finally we also include additional saturation scale fluctuations for each constituent quark separately following Ref.~\cite{McLerran:2015qxa}, where it is shown that proton-proton multiplicity fluctuations can be described within the IP-Glasma framework if $\ln Q_s^2/\langle Q_s^2\rangle$ fluctuates according to a Gaussian distribution whose width is $\sigma \approx 0.5$. We find that the saturation scale fluctuations mainly affect the incoherent cross section at the smallest $|t|$, and that for larger momentum transfer the dominant contribution originates from the geometric fluctuations. This result is expected based on Ref.~\cite{Miettinen:1978jb}.

\section{Conclusions}
We have shown how diffractive vector meson production measured in deep inelastic scattering experiments at HERA can be used to constrain the amount of event-by-event fluctuations of the proton density. We find that significant geometric fluctuations are needed in order to obtain an incoherent cross section compatible with the experimental data. The color charge fluctuations, on the other hand, are found to have small effect on the incoherent cross section. Similarly, the effect of additional saturation scale fluctuations on the incoherent cross section is dominant only at small $|t|$.
The inclusion of these results into the initial state for hydrodynamical calculations 
of  proton-nucleus collisions is currently underway.




\bibliographystyle{elsarticle-num}






\bibliography{../../refs}

\begin{thebibliography}{10}
\expandafter\ifx\csname url\endcsname\relax
  \def\url#1{\texttt{#1}}\fi
\expandafter\ifx\csname urlprefix\endcsname\relax\def\urlprefix{URL }\fi
\expandafter\ifx\csname href\endcsname\relax
  \def\href#1#2{#2} \def\path#1{#1}\fi

\bibitem{Abramowicz:2015mha}
H.~Abramowicz, et~al., {Combination of measurements of inclusive deep inelastic
  ${e^{\pm }p}$ scattering cross sections and QCD analysis of HERA data}, Eur.
  Phys. J. C75 (2015) 580.
\newblock \href {http://arxiv.org/abs/1506.06042} {\path{arXiv:1506.06042}},
  \href {http://dx.doi.org/10.1140/epjc/s10052-015-3710-4}
  {\path{doi:10.1140/epjc/s10052-015-3710-4}}.

\bibitem{Belitsky:2003nz}
A.~V. Belitsky, X.-d. Ji, F.~Yuan, {Quark imaging in the proton via quantum
  phase space distributions}, Phys. Rev. D69 (2004) 074014.
\newblock \href {http://arxiv.org/abs/hep-ph/0307383}
  {\path{arXiv:hep-ph/0307383}}, \href
  {http://dx.doi.org/10.1103/PhysRevD.69.074014}
  {\path{doi:10.1103/PhysRevD.69.074014}}.

\bibitem{Accardi:2012qut}
A.~Accardi, et~al., {Electron Ion Collider: The Next QCD Frontier -
  Understanding the glue that binds us all}, Eur. Phys. J. A52 (2016) 268.
\newblock \href {http://arxiv.org/abs/1212.1701} {\path{arXiv:1212.1701}},
  \href {http://dx.doi.org/10.1140/epja/i2016-16268-9}
  {\path{doi:10.1140/epja/i2016-16268-9}}.

\bibitem{Mantysaari:2016ykx}
H.~Mäntysaari, B.~Schenke, {Evidence of strong proton shape fluctuations from
  incoherent diffraction}, Phys. Rev. Lett. 117 (2016) 052301.
\newblock \href {http://arxiv.org/abs/1603.04349} {\path{arXiv:1603.04349}},
  \href {http://dx.doi.org/10.1103/PhysRevLett.117.052301}
  {\path{doi:10.1103/PhysRevLett.117.052301}}.

\bibitem{Mantysaari:2016jaz}
H.~Mäntysaari, B.~Schenke, {Revealing proton shape fluctuations with
  incoherent diffraction at high energy}, Phys. Rev. D94~(3) (2016) 034042.
\newblock \href {http://arxiv.org/abs/1607.01711} {\path{arXiv:1607.01711}},
  \href {http://dx.doi.org/10.1103/PhysRevD.94.034042}
  {\path{doi:10.1103/PhysRevD.94.034042}}.

\bibitem{Dusling:2015gta}
K.~Dusling, W.~Li, B.~Schenke, {Novel collective phenomena in high-energy
  proton–proton and proton–nucleus collisions}, Int. J. Mod. Phys. E25
  (2016) 1630002.
\newblock \href {http://arxiv.org/abs/1509.07939} {\path{arXiv:1509.07939}},
  \href {http://dx.doi.org/10.1142/S0218301316300022}
  {\path{doi:10.1142/S0218301316300022}}.

\bibitem{Good:1960ba}
M.~L. Good, W.~D. Walker, {Diffraction disssociation of beam particles}, Phys.
  Rev. 120 (1960) 1857--1860.
\newblock \href {http://dx.doi.org/10.1103/PhysRev.120.1857}
  {\path{doi:10.1103/PhysRev.120.1857}}.

\bibitem{Miettinen:1978jb}
H.~I. Miettinen, J.~Pumplin, {Diffraction Scattering and the Parton Structure
  of Hadrons}, Phys. Rev. D18 (1978) 1696.
\newblock \href {http://dx.doi.org/10.1103/PhysRevD.18.1696}
  {\path{doi:10.1103/PhysRevD.18.1696}}.

\bibitem{Frankfurt:1993qi}
L.~Frankfurt, G.~A. Miller, M.~Strikman, {Evidence for color fluctuations in
  hadrons from coherent nuclear diffraction}, Phys. Rev. Lett. 71 (1993)
  2859--2862.
\newblock \href {http://arxiv.org/abs/hep-ph/9309285}
  {\path{arXiv:hep-ph/9309285}}, \href
  {http://dx.doi.org/10.1103/PhysRevLett.71.2859}
  {\path{doi:10.1103/PhysRevLett.71.2859}}.

\bibitem{Caldwell:2009ke}
A.~Caldwell, H.~Kowalski, {Investigating the gluonic structure of nuclei via
  $J/\Psi$ scattering}, Phys. Rev. C81 (2010) 025203.
\newblock \href {http://arxiv.org/abs/0909.1254} {\path{arXiv:0909.1254}},
  \href {http://dx.doi.org/10.1103/PhysRevC.81.025203}
  {\path{doi:10.1103/PhysRevC.81.025203}}.

\bibitem{Kowalski:2006hc}
H.~Kowalski, L.~Motyka, G.~Watt, {Exclusive diffractive processes at HERA
  within the dipole picture}, Phys. Rev. D74 (2006) 074016.
\newblock \href {http://arxiv.org/abs/hep-ph/0606272}
  {\path{arXiv:hep-ph/0606272}}, \href
  {http://dx.doi.org/10.1103/PhysRevD.74.074016}
  {\path{doi:10.1103/PhysRevD.74.074016}}.

\bibitem{Albacete:2010sy}
J.~L. Albacete, N.~Armesto, J.~G. Milhano, P.~Quiroga-Arias, C.~A. Salgado,
  {AAMQS: A non-linear QCD analysis of new HERA data at small-x including heavy
  quarks}, Eur. Phys. J. C71 (2011) 1705.
\newblock \href {http://arxiv.org/abs/1012.4408} {\path{arXiv:1012.4408}},
  \href {http://dx.doi.org/10.1140/epjc/s10052-011-1705-3}
  {\path{doi:10.1140/epjc/s10052-011-1705-3}}.

\bibitem{Lappi:2013zma}
T.~Lappi, H.~M{\"a}ntysaari, {Single inclusive particle production at high
  energy from HERA data to proton-nucleus collisions}, Phys. Rev. D88 (2013)
  114020.
\newblock \href {http://arxiv.org/abs/1309.6963} {\path{arXiv:1309.6963}},
  \href {http://dx.doi.org/10.1103/PhysRevD.88.114020}
  {\path{doi:10.1103/PhysRevD.88.114020}}.

\bibitem{Rezaeian:2012ji}
A.~H. Rezaeian, M.~Siddikov, M.~Van~de Klundert, R.~Venugopalan, {Analysis of
  combined HERA data in the Impact-Parameter dependent Saturation model}, Phys.
  Rev. D87 (2013) 034002.
\newblock \href {http://arxiv.org/abs/1212.2974} {\path{arXiv:1212.2974}},
  \href {http://dx.doi.org/10.1103/PhysRevD.87.034002}
  {\path{doi:10.1103/PhysRevD.87.034002}}.

\bibitem{Lappi:2010dd}
T.~Lappi, H.~M{\"a}ntysaari, {Incoherent diffractive $J/\Psi$-production in
  high energy nuclear DIS}, Phys. Rev. C83 (2011) 065202.
\newblock \href {http://arxiv.org/abs/1011.1988} {\path{arXiv:1011.1988}},
  \href {http://dx.doi.org/10.1103/PhysRevC.83.065202}
  {\path{doi:10.1103/PhysRevC.83.065202}}.

\bibitem{Lappi:2013am}
T.~Lappi, H.~M{\"a}ntysaari, {$J/\Psi$ production in ultraperipheral Pb+Pb and
  p+Pb collisions at LHC energies}, Phys. Rev. C87 (2013) 032201.
\newblock \href {http://arxiv.org/abs/1301.4095} {\path{arXiv:1301.4095}},
  \href {http://dx.doi.org/10.1103/PhysRevC.87.032201}
  {\path{doi:10.1103/PhysRevC.87.032201}}.

\bibitem{Schenke:2012fw}
B.~Schenke, P.~Tribedy, R.~Venugopalan, {Event-by-event gluon multiplicity,
  energy density, and eccentricities in ultrarelativistic heavy-ion
  collisions}, Phys. Rev. C86 (2012) 034908.
\newblock \href {http://arxiv.org/abs/1206.6805} {\path{arXiv:1206.6805}},
  \href {http://dx.doi.org/10.1103/PhysRevC.86.034908}
  {\path{doi:10.1103/PhysRevC.86.034908}}.

\bibitem{Aaron:2009xp}
F.~Aaron, et~al., {Diffractive Electroproduction of $\rho$ and $\phi$ Mesons at
  HERA}, JHEP 1005 (2010) 032.
\newblock \href {http://arxiv.org/abs/0910.5831} {\path{arXiv:0910.5831}},
  \href {http://dx.doi.org/10.1007/JHEP05(2010)032}
  {\path{doi:10.1007/JHEP05(2010)032}}.

\bibitem{McLerran:2015qxa}
L.~McLerran, P.~Tribedy, {Intrinsic Fluctuations of the Proton Saturation
  Momentum Scale in High Multiplicity p+p Collisions}, Nucl. Phys. A945 (2016)
  216--225.
\newblock \href {http://arxiv.org/abs/1508.03292} {\path{arXiv:1508.03292}},
  \href {http://dx.doi.org/10.1016/j.nuclphysa.2015.10.008}
  {\path{doi:10.1016/j.nuclphysa.2015.10.008}}.

\end{thebibliography}

\subsection*{Acknowledgments}
This work was supported under DOE Contract No. DE-SC0012704 and used resources of the National Energy Research Scientific Computing Center, supported by the Office of Science of the U.S. Department of Energy under Contract No. DE-AC02-05CH11231. BPS acknowledges a DOE Office of Science Early Career Award.

\end{document}